# Zero Bias Power Detector Circuits based on MoS$_2$ Field Effect Transistors on Wafer-Scale Flexible Substrates

*Eros Reato[1,2], Paula Palacios[3], Burkay Uzlu[1,2], Mohamed Saeed[3], Annika Grundmann[4], Zhenyu Wang[5], Daniel S. Schneider[1], Zhenxing Wang[1,\*], Michael Heuken[4,6], Holger Kalisch[4], Andrei Vescan[4], Alexandra Radenovic[5], Andras Kis[5], Daniel Neumaier,[1,7] Renato Negra[3,\*], Max C. Lemme[1,2,\*]*

[1] AMO GmbH, Advanced Microelectroncis Center Aachen (AMICA), Otto-Blumenthal-Strasse 25, 52074 Aachen, Germany

[2] Chair of Electronic Devices, RWTH Aachen University, Otto-Blumenthal-Strasse 25, 52074 Aachen, Germany

[3] Chair of High Frequency Electronics, RWTH-Aachen University, Kopernikusstraße 16, 52074 Aachen, Germany

[4] Compound Semiconductor Technology, RWTH Aachen University, Sommerfeldstrasse 18, 52074 Aachen, Germany

[5] School of Engineering, EPFL, BM 2141, Station 17, 1015 Lausanne, Switzerland

[6] AIXTRON SE, Dornkaulstrasse 2, 52134 Herzogenrath, Germany

[7] Bergische Universität Wuppertal, Lise-Meitner-Str. 13, 42119 Wuppertal

Corresponding authors: wang@amo.de, negra@hfe.rwth-aachen.de, max.lemme@eld.rwth-aachen.de

Eros Reato and Paula Palacios contributed equally to this work.

**Abstract**

We demonstrate the design, fabrication, and characterization of wafer-scale, zero-bias power detectors based on two-dimensional $MoS_2$ field effect transistors (FETs). The $MoS_2$ FETs are fabricated using a wafer-scale process on 8 μm thick Polyimide film, which in principle serves as flexible substrate. The performances of two CVD-$MoS_2$ sheets, grown with different processes and showing different thicknesses, are analyzed and compared from the single device fabrication and characterization steps to the circuit level. The power detector prototypes exploit the nonlinearity of the transistors above the cut-off frequency of the devices. The proposed detectors are designed employing a transistor model based on measurement results. The fabricated circuits operate in *Ku*-band between 12 and 18 GHz, with a demonstrated voltage responsivity of 45 V/W at 18 GHz in the case of monolayer $MoS_2$ and 104 V/W at 16 GHz in the case of multilayer $MoS_2$, both achieved without applied DC bias. They are the best performing power detectors fabricated on flexible substrate reported to date. The measured dynamic range exceeds 30 dB outperforming other semiconductor technologies like silicon complementary metal oxide semiconductor (CMOS) circuits and GaAs Schottky diodes.

1. **Introduction**

The mechanical flexibility and electronic transport properties of two-dimensional (2D) materials allow their integration on flexible substrates and provide a high potential for transparent bendable and wearable electronics. With a higher charge carrier mobility than organic semiconductors, 2D materials, such as graphene, black phosphorus (BP) and transition metal dichalcogenides (TMDs), are in the lead towards the realization of such applications that cannot be achieved with conventional semiconductor technologies[1–8]. Pioneered by the successful mechanical exfoliation of graphene in 2004[9], an ample number of articles have been published with graphene-based devices and circuits, both on rigid and flexible substrates[10–18]. However, the lack of a bandgap in graphene intrinsically prevents graphene field effect transistors (FETs) from achieving a distinct OFF state in digital circuitry and limits their achievable current saturation in the output characteristics (drain current, $I_{ds}$, versus source-drain voltage, $V_{ds}$). The latter considerably lowers $f_{max}$ of the transistors and limits their performance in terms of power gain in high-frequency designs[19–22]. In contrast to graphene, BP possesses a thickness-dependent direct bandgap that enables transistors with a high ON/OFF current ratio[23,24]. BPs appealing characteristics are somewhat compromised by the lack of stability under ambient conditions and a resulting rapid degradation of its electrical properties[7], although encapsulation methods may mitigate this fact[25,26]. Molybdenum disulphide ($MoS_2$) has attracted interest as a semiconducting 2D material with high stability in ambient air[1]. The presence of an energy gap between 1.3 eV and 1.8 eV in the multi- and single-layer form permits ON/OFF ratios as high as $10^8$ in $MoS_2$ FETs. This enables electronic circuits, both in the digital[27,28] and the relatively high frequency analogue[29] domains. In addition, the current saturation in the device output characteristics have led to maximum extrinsic $f_T$ and $f_{max}$ of 4 GHz and 10 GHz, respectively, in devices on flexible substrates [4]. Finally, $MoS_2$-based diodes have been used to build low power flexible integrated transceivers[30,31].

Based on these considerations, $MoS_2$-based low-power wireless transceivers are an attractive and feasible goal towards implementing high performance microwave electronic circuits on flexible substrates. However, while RF mixers[5,32] and amplifiers[6] operating in the megahertz range have been reported, power detectors as important building blocks are still missing. They are typically employed either to detect small signals close to the noise level or to monitor large signal levels. The power detector relies on the non-linear operation of a single FET. Proper gate biasing (in this case 0 V) can be applied to operate the device in the square-law region to obtain a DC voltage/current at the transistor drain which is proportional to the RF power delivered to the gate. The working principle relies on the nonlinear characteristic of the active device(s) at the operating point. Under a small-signal voltage, $av,$ the current-voltage relation of the device can be represented by a Taylor expansion[33]

$$I = f(V) = f(V_0) + \frac{df}{dV}(av) + \frac{1}{2}\frac{d^2f}{d^2V}(av)^2 \qquad (1)$$

where $V_0$ is the bias voltage that defines the operation point and where the derivatives are evaluated. By assuming that higher order terms are sufficiently small, and due to the symmetry of the linear term, the DC component originates from the second-order term. Therefore, detectors function as square-law rectifiers when the input signal is sufficiently small. Graphene p-n junction or Schottky diodes have been used as power detectors due to their high performance together with their well-established process technology[34,15]. However, these are not easily integrable on-chip and most designs require biasing. Consequently, the noise performance is lowered in comparison to zero-bias designs, which only exhibit thermal noise. Moreover, silicon CMOS-based designs have also been implemented as integrated power detectors[35,36], but are limited in dynamic range and are not suitable for flexible substrates.

Power detectors are generally used in numerous analog wireless applications in different fields such as radar systems, Radio Frequency Identification (RFID) transceivers, or mobile communications[37]. In addition, they are one of the basic components in six-port receivers, together with local oscillators and low-noise amplifiers (LNA)[8]. The architectures of such receivers present lower complexity in comparison to other front-ends, and potentially allows fully integrated flexible RF front-end design when based on $MoS_2$ technology. In this work, we report the successful implementation of zero-bias RF power detectors based on two different $MoS_2$ FETs with mono- and multilayer channel materials, both fabricated with a scalable and manufacturable growth technique.

## 2. MoS$_2$ FET fabrication and characterization

The $MoS_2$ films were grown by metal-organic vapor phase deposition (MOCVD) [38,39] on 2-inch sapphire wafers. Multilayer material (M) was deposited using di-tert-butyl-sulfide (DTBT) as the sulfur precursor[38], while $H_2S$ was selected for the monolayer material (S) deposition[39]. The Raman spectra of both materials after transfer onto $SiO_2$ test substrates show that the distance between the $E^1_{2g}$ and $A_{2g}$ peaks of the M-material is larger than for the S-material, indicating a larger thickness of the former[40] **(Figure 1a)**. The statistical distributions of this peak separation, extracted with the help of 625 Raman spectroscopic measurements in an area of 50 μm x 50 μm, are shown in the inset of **Figure 1a**. The difference in thickness of the two materials, as well as the surface roughness and morphology, was confirmed by AFM scans, the single layer material (S) is 0.9 nm thick while the multilayer material is (M) is 5 nm thick. More information on the AFM scans can be found in Section 1 of the Supporting Information.

The $MoS_2$ FETs were fabricated on a flexible 8 μm thick polyimide (PI) film on silicon wafers with standard photolithography. The flexible layer of the desired thickness was obtained by spincoating the liquid PI on a Si carrier substrate and by curing the substrate at 350 °C for 30 minutes. The back gate consists of a 100 nm thick aluminum (Al) layer followed by a 35 nm thick titanium (Ti) layer, which was deposited via electron beam evaporation. An aluminum oxide ($Al_2O_3$) gate insulator with a thickness of 35 nm was deposited in an atomic layer deposition (ALD) system. The oxide thicknesses were confirmed by ellipsometry. Commercially available hexagonal boron nitride (h-BN) layers and different $MoS_2$ layers were transferred from their respective growth substrates onto these prepatterned PI on silicon substrates with a wet transfer technique. The $MoS_2$ was then covered with a second layer of h-BN in order to improve the quality of the interface[41,42]. The channel was patterned by reactive ion etching (RIE) and the electrical contacts to the h-BN/$MoS_2$/h-BN stack were realized by DC sputtering of a 50 nm thick nickel layer. The DC sputtering, combined with the RIE step allowed us to contact the $MoS_2$ layer from the edge of the material, thus creating edge contacted devices. This contact scheme is results in reasonable values of contact resistance [43–45] although more recent methods have emerged that we have not yet been able to implement.[46] After the contacts, a 75 nm thick encapsulation layer of $Al_2O_3$ was deposited by ALD. Finally, electron beam evaporation was used again to deposit the final Al metallization as probing pads. A schematic of the devices is shown in **Figure 1b** and optical micrographs of mono- and multilayer MoS$_2$ devices with two parallel channels are shown in **Figure 1c** and **d,** respectively. The mask layout for the MoS$_2$ FETs includes different channel dimensions and the same mask set was used for the M- and the S-type devices. DC transfer (drain current, $I_{ds}$, versus gate voltage, $V_{gs}$, on **Figure 2a**) and output (drain current, $I_{ds}$, versus drain voltage, $V_{ds}$, on **Figure 2b** and **c**) characteristics were measured at room temperature and ambient air with a semiconductor parameter analyzer. The gate length and width of these two MoS$_2$ FETs are $L = 6$ µm and $W = 60$ µm, respectively. The transfer characteristics were measured at a constant $V_{ds} = 1$ V with increasing and decreasing $V_{gs}$ sweeps, plotted as solid and dashed lines, respectively. An ON current of 12.8 µA and 16.8 µA was measured for the M- and S-materials, respectively. We define ON current as the value of current flowing across the device for a specific $V_{gs} - V_{th} > 0$ V (in this case $V_{gs} - V_{th} = 4.4$ V), *i.e.* when the transistor is biased above the threshold voltage, $V_{TH}$. The transfer characteristic allowed us to extract the carrier mobility from the transconductance ($gm$) of the devices. The resulting mobility is $\mu_{gm,M} = 1.5$ cm$^2$/V·s and $\mu_{gm,S} = 1.8$ cm$^2$/V·s for the multilayer and monolayer materials, respectively. The analysis of the curves with the Y function method ($Ids/\sqrt{gm}$) allowed the calculation of the low field mobility ($\mu 0$) and contact resistance ($R_C$) from a single transfer curve, and resulted in $\mu_{0,M} = 1.63$ cm$^2$/V·s, $\mu_{0,S} = 2.18$ cm$^2$/V·s, $R_{C,S} = 100$ kΩ·µm and $R_{C,M} = 131$ kΩ·µm. It is known that the low field mobility from the Y function is an extraction method that excludes the effects of the contact resistance, therefore higher $\mu 0$ values were expected[47,48]. The constant current method was used to estimate the value of the threshold voltage[49]. For a defined drain current of 1 µA, we obtain $V_{TH, M} = -7.5$ V and $V_{TH, S} = -4.4$ V. A small clockwise hysteresis is observed for the M devices, which is a common phenomenon in MOSFETs with MoS$_2$ channels. It is usually an indicator of the presence of border traps which are negatively charged during the sweep in the gate oxide of the transistors[41,50]. One can observe some substantial differences in the curves other than hysteresis for the M-material, for example a clear degradation of the inverse Subthreshold Slope (or Subthreshold Swing, SS) and a reduction of the ON/OFF ratio, compared to the S-material. This may be attributed to the presence of a higher amount of defects, such as sulfur vacancies, in the multilayer material.[51] Since sulfur vacancies have been found to be responsible for an n-doping effect of $MoS_2$[52], this would be also consistent with the larger negative threshold voltage of the M-device. Although many methods for the reduction of hysteresis and defect passivation exist, they usually require either high temperature annealing steps [53,54] or non-cleanroom-standard chemical treatments[55,56] and they are not used in this work due to the lack of comparative studies in literature about such processes for flexible devices. The devices were additionally tested on bending rods with different radii and after several bending cycles. Both devices batches show some degradation of the performances after the peeling of the devices. The devices fabricated with the S-material gets damaged after the bending tests with the minimum bending radius, while the devices fabricated with the M-material are able to withstand up to 1000 bending cycles without major performances loss. The results are summarized in the Supporting Information while a compilation of the main DC characteristics of both devices batches is summarized in **Table 1**. The high frequency response was characterized by standard two-port *S*-parameter measurements with different bias conditions ranging from -16 V to 5 V for $V_{gs}$ in the M-transistors, -7 V to 5 V for $V_{gs}$ in the S-transistors, and from 0 V to 11 V for $V_{ds}$. The channel lengths of the RF measured S- and M-transistors were 6 μm and 5 μm, respectively, while the width was 60 μm for both devices. Measurements were carried out from 10 MHz to 40 GHz on-wafer by means of ground-signal-ground (GSG) pads attached to the intrinsic transistor, as shown in **Figure 1c**. The resulting $S_{11}$, $S_{22}$, and $S_{21}$ are plotted in **Figure 3** for 0 V at the gate and drain. As expected for symmetric devices, $S_{12}$ and $S_{21}$ are equal. From the *S*-parameter measurements, the current gain, $h_{21}$, and the maximum available gain (MAG) were calculated in order to extract the transient frequency, $f_T$, and the maximum oscillation frequency, $f_{max}$, (**Figure 4**). Defined as the magnitude at which $h_{21}$ becomes unity, $f_T$ equals 57.7 MHz and 33.27 MHz for the M- and S-transistors, respectively. Similarly, defined as the magnitude at which the maximum available gain (MAG) equals unity, the obtained $f_{max}$ is 236.6 MHz and 114.1 MHz. These figures-of-merit were extracted in both transistor types for a $V_{gs}$ = 1 V and a $V_{ds}$ = 11 V. The performance is in line with typical values of state-of-the-art flexible MoS$_2$-based RF devices[6,32] considering the dimensions of the transistors. $f_T$ and $f_{max}$ of a MOSFET, considering drift-diffusion transport and without considering short-channel effects, can be expressed as

$$f_T = \frac{1}{2\pi}\frac{g_m}{C_{gs}} = \frac{1}{2\pi}\frac{3}{2}\frac{\mu}{L^2}(V_{gs} - V_{th}) \qquad (2)$$

and

$$f_{max} = \frac{1}{2}\frac{f_T}{\sqrt{R_g(g_{ds} + f_T C_{gs})}} \qquad (3),$$

where $\mu$ is the carrier mobility, $L$ is the gate length, $R_g$ is the gate electrode resistance, and $g_{ds}$ is the output conductance[57]. $f_T$ is inversely proportional to the square of the length of the channel (equ. 2), which explains the difference in performance between the S- and M-devices.

Although there is great interest in improving these two FOMs of transistors, only microwave circuit applications that make use of the transconductance rely on them[58]. When operated in the linear region at zero $V_{ds}$ bias, transistors behave as a nonlinear gate voltage-controlled resistor and allow applications beyond $f_T$ and $f_{max}$. Since the mobility also relies on the transconductance, the same argument can be used to explain the high performance of the circuit in terms of frequency (section 3), despite the extracted low mobility values (**Table 1**). The extracted model from *S*-parameter measurements at this operating point ($V_{ds}$ = 0V) can be simplified as shown in **Figure 5a,** where no $g_m$ is included. The values of the lumped components of the transistor model were extracted from *S*-parameter measurements according to reference [59] and are listed in

**Table 2.** The equivalent circuit encompasses the intrinsic transistor at the operating point and the parasitic elements of the gate, source, and drain series resistances. Moreover, the resistance to the substrate is included as well as an extra capacitance at the source that increases the asymmetry between the ports. These are related to the pads and the measurement parasitics. Simulations of the equivalent circuit for the transistor at zero bias were performed with the Advanced Design Systems (ADS) software from Keysight, and a good agreement between the model and the measurement results was achieved (**Figure 3**).

## 3. MoS$_2$ Power detector circuit

Power detection requires a nonlinearity in the circuit, as presented in equation (1). In the proposed design, this is incorporated through the modulation of the channel resistance with the gate voltage. Represented as $R_{DS}$ in Fehler! Verweisquelle konnte nicht gefunden werden.**a,** the resistance variation with respect to $V_{gs}$ is modeled from the S-parameters for different gate bias voltages and constant $V_{ds} = 0$ V. This nonlinear response of $R_{DS}$ for the two transistor types can be observed close to their respective threshold voltage in Fehler! Verweisquelle konnte nicht gefunden werden.**b**. Effectively, the transistors can be considered in their OFF-states at $V_{TH}$. This behavior was implemented in the simulation with a high order polynomial fitted to the extracted values. The difference between the $V_{TH}$ values obtained in the DC characterization and the ones presented in the model could be ascribed to some instability in the threshold voltage values. The analysis of the threshold voltage stability measurements is summarized in the Supporting Information Section 3.

The power detector architecture is depicted in **Figure 6a,** where the transistors are on-wafer and further elements are connected externally. The input signal with calibrated power, $P_{in}$, is provided by the RF source of the microwave vector network analyzer (VNA) through the internal bias tee. The VNA is connected to the gate of the transistor and the bias voltage is set to 0 V. At the gate input an external 50-Ω coaxial load is included to match the input. The output signal is conducted from the drain to an external bias tee that acts as a lowpass filter (LPF). The RF signal is short-circuited to ground and the DC output is measured with the voltmeter across an 800-kΩ load. Since the expected frequency performance is not related to the $f_T$ and $f_{max}$ of the transistor, the 3-dB RF detection bandwidth is defined as[60]

$$f_{3-dB} = 1/\left(2\pi(C_{gs} + C_{gd})R_g\right) \qquad (4).$$

This frequency is calculated from the equivalent circuit to be approximately 122 GHz and 128 GHz for the M- and S-transistors, respectively, far beyond the cutoff frequencies of the transistors. Nevertheless, this estimated value considers neither the parasitic components nor the nonideal externally connected components.

Previous to the actual power detector measurements, one-port *S*-parameter measurements were carried out in order to identify the best matched frequencies up to 50 GHz. With a -10-dB matching criteria, $S_{11}$ at the input of the VNA was used as the determining factor to choose the frequencies for the power detection setup. Based on these measurements, 16 GHz and 18 GHz were chosen for the M-and S-FETs, respectively. The measured output voltage as a function of the input RF-power for these two frequency responses are shown in **Figure 6b.** Both transistor types present a similar performance in terms of dynamic range, where the linear-in-dB region extends from -30 dBm up to 0 dBm of input power. The large dynamic range of 30 dB is comparable to graphene diode-based power detectors and other FET-based circuits fabricated on rigid substrates[14,60–62]. The responsivity of the power detector circuits is defined as

$$Responsivity = V_{out}/P_{in} \ [V/W] \qquad (5),$$

where $V_{out}$ is the output DC voltage and $P_{in}$ is the input RF power. The M-transistors show a slightly better performance than the S-devices. Extracted from the slope in **Figure 6b,** the responsivity is presented in **Figure 6c.** The highest responses are at 16 GHz and 11 GHz in the M- and 18 GHz and 21 GHz in the S-devices, with measured responsivities of 104 V/W and 134 V/W, as well as 45 V/W and 49 V/W, respectively.

Some fluctuations can be observed in the responsivity plots (**Figure 6c**). Moreover, the theoretical maximum $f_{3-dB}$ was not achieved in measurements. This is (in part) due to the external components in the setup. Since the circuit relies on a nonideal resistive load for matching and a bias tee at the output, the varying insertion loss over the working frequency range of these devices affects the overall performance of the detector in different ways. The lack of matching prevents the input power from reaching the transistors, and the reflection at the bias tee may lead to spurious higher-order harmonics that lower the conversion efficiency. In order to verify this behavior and to validate the nonlinear resistance model implemented in **Figure 5a** a harmonic balance simulation of the full power detector was performed, where the active device was substituted by the extracted equivalent circuit. The *S*-parameters of the 50-Ω load and of the output bias tee were measured and included in the simulation for higher accuracy. Moreover, two extra inductors were added at the output of both the VNA and of the circuit model to include additional inductance originating from the cables. The resulting comparison between simulations and measurements is shown in **Figure 7a** and **b,** where the output voltage is plotted over the input power for different frequencies. In agreement with the experimental results, certain frequencies present higher responsivities that are related to the matching in the circuit also in simulations. Simulations and measurements are in good agreement, and deviations are direct consequences of shifts in the matched frequencies caused by the inevitable discrepancies between the *S*-parameters used in simulation and the real external components, cables, and adaptors. Moreover, since the real components were included as *S*-parameters measured up to 40 GHz, higher harmonics were considered by interpolation in simulation. It is important to highlight that the model does not include the noise level, thus, it is not useful to determine the dynamic range of the detector. This is reflected in the simulations at low power levels in **Figure 7a** and **b,** where the model continues its linear behavior and deviates clearly from the experiments. Furthermore, the detectors are driven into saturation at high input power levels. To capture this effect in simulations, a large-signal model of the transistor would be required. These facts indicate that in addition to the successful demonstration of rectification at such high frequencies, even higher performance can be expected from a fully integrated solution with application-customized device structures, since this would avoid all the mismatches and tolerances from the external devices.

A comparison with the state-of-the-art power detectors built in different technologies based on, both, bulk semiconductors and 2D materials is summarized in **Table 3**, where all the selected designs are implemented using a single device. In [63], the circuit is implemented in 130 nm CMOS technology, and whereas it presents a great responsivity, it is at the expense of a high

DC power consumption and a limited bandwidth. On the other hand, [64] and [65], are based on two different types of GaAs diodes, present excellent responsivities and higher bandwidth, but they are limited in dynamic range due to their high junction resistance and being cost inefficient. Furthermore, the three so far mentioned technologies do not allow their integration on flexible substrates. Then, graphene-based power detectors, reported in [60] and [15], based on a GFET and a MIG diode, respectively, have proved a higher dynamic range than standard technologies for a wide bandwidth at zero bias. Nevertheless, both designs, contrarily to our reported design, are based on rigid substrates and show lower responsivities. Therefore, this work demonstrates the potential of MoS2 FETs as power detectors for flexible electronics, outperforming the responsivities of other single-device hybrid designs based on 2D materials.

Finally, the M-material-based circuit was incorporated into a real system and characterized as an ON-OFF keying (OOK) signal demodulator at 16.8 GHz. This practical application of the power detector is shown in **Figure 8a**. The OOK modulated baseband (BB) signal was generated by the signal generator and upconverted to the 16 GHz band by a passive mixer. A local oscillator (LO) was provided by the vector network analyzer, whose maximum available power is 0 dBm. In order to compensate the power losses from the mixer, cables, and connectors, a 30-dB amplifier was used at the output of the mixer to guarantee a sufficient signal level at the input of the $MoS_2$ transistor. The voltage signal was measured by an oscillator with a 1-MΩ input impedance. The obtained waveforms at 1 kSymb/s and 100 kSymb/s modulation rates are shown in Fehler! Verweisquelle konnte nicht gefunden werden.**b** and **Figure 8c**, respectively. The high responsivity of the power detector is demonstrated in the peak-to-peak voltage, equal to over 100 mV for the lowest rate. At 1 kSymb/s the ON and OFF-states are clearly distinguishable (Fehler! Verweisquelle konnte nicht gefunden werden.**b)**. At 100 kSymb/s, the symbols are still discernible, although the waveform is deformed, and the signal does not decay sufficiently fast when an ON-state immediately follows an OFF-state. This is because the circuit was optimized for its responsivity and linearity rather than demodulation. Therefore, the output bias tee together with the input impedance of the oscilloscope filter the signal and prevent higher modulation rates. Nevertheless, rates as high as 100 kSymb/s can be demodulated, and by incorporating an amplifier and comparator after the detector, the information can be digitally processed in a receiver.

**4. Conclusions**

$MoS_2$-based devices allow the monolithic integration of digital and analogue circuits on the same flexible substrate. Nevertheless, $MoS_2$ FET-based microwave applications are limited by the performance of the transistors in terms of their $f_T$ and $f_{max}$. In this work, zero-bias power detectors based on $MoS_2$ FETs are implemented for the first time. By relying on the nonlinearity of the channel, the operation frequency is far above the cutoff frequency reported for flexible devices up to date. The power detectors show also high performance in terms of dynamic range and responsivity and are in good agreement with simulation results, with the highest performances among the flexible technologies. It has been established that despite the good performance of other semiconductor technologies, $MoS_2$ allows the implementation of wide dynamic range power detectors with zero DC power consumption on flexible substrates. Furthermore, in contrast to other 2D material-based technologies, $MoS_2$ transistors are stable at room temperatures and do present a high ON/OFF ratio allowing the integration of full flexible transceivers. In addition, a circuit has been proven capable of successfully demodulating OOK signals with up 100 kSymb/s rates at 16.8 GHz, ultimately demonstrating its potential application when incorporated into a real system. However, this work also shows that still some material growth and device processing parameters need to be carefully tuned in order to allow these circuits to meet the rigorous reliability requirements for mass production. Nonetheless, this work show that the promising performance, the zero-power consumption and the fabrication on a flexible substrate pave the way for exploiting this technology in the future bendable and wearable low-power electronic devices.

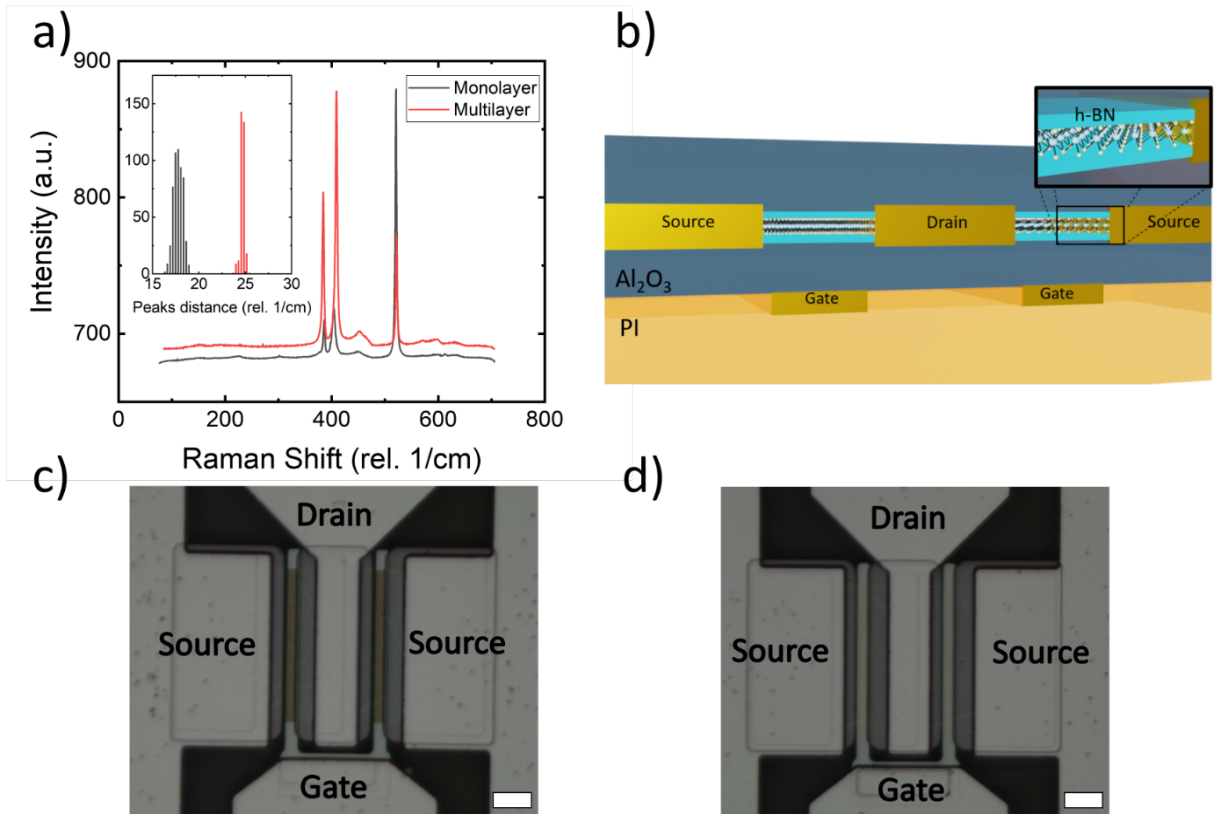

**Figure 1** a) Raman characterization of both MoS$_2$ batches after the transfer onto a SiO$_2$ test substrate. The inset shows the statistical distribution of the distance between the E$^1_{2g}$ and A$_{2g}$ resonance peaks of MoS$_2$. b) Schematic cross-section of the MoS$_2$ devices. c) Optical micrograph of the multilayer MoS$_2$ device. d) Optical micrograph of the monolayer MoS$_2$ device. The scale bar is 16 μm.

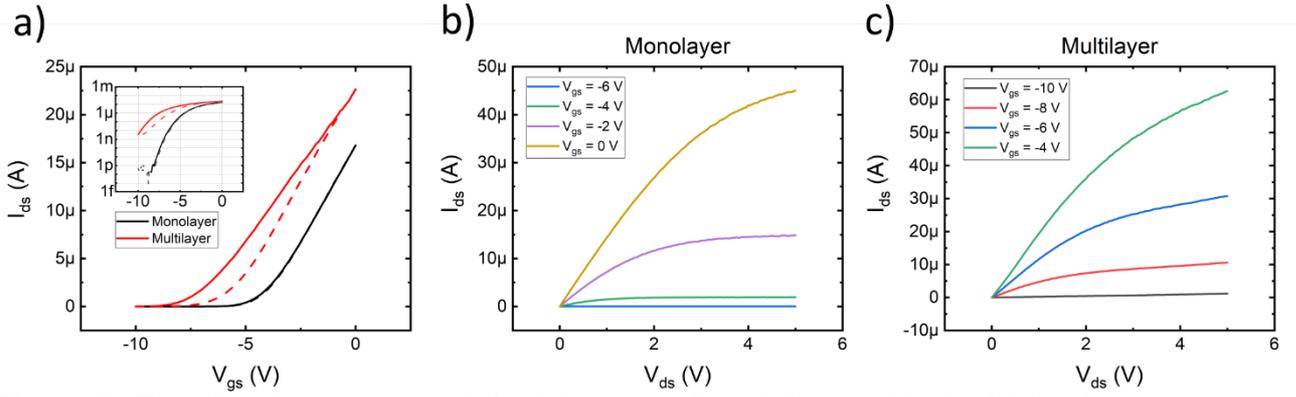

**Figure 2.** Transfer characteristic of the M-type (red) and S-type (black) MoS$_2$ devices at $V_{ds} = 1$ V. The dimensions of the devices are $L = 6$ μm and $W = 60$ μm. The dashed lines represent backward voltage sweep and show the presence of hysteresis in the measurements. In the inset the same data is plotted in a semilog scale. b) Output characteristic of an monolayer MoS$_2$ device. c) Output characteristic of a multilayer MoS$_2$ device.

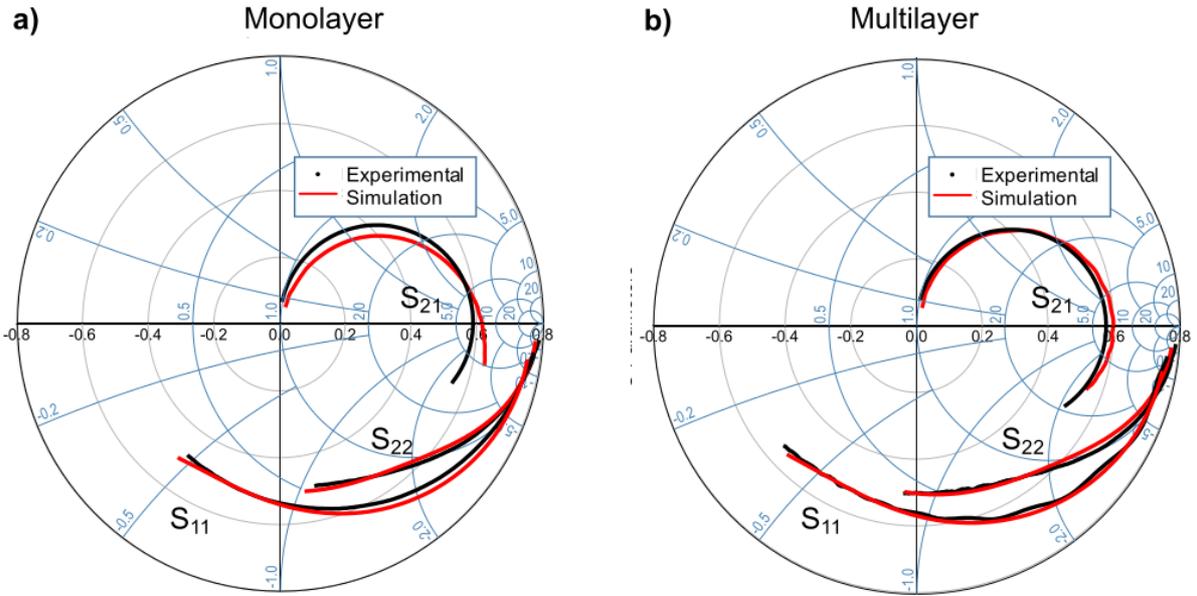

**Figure 3.** *S*-parameter measurement results at 0 V $V_{gs}$ and 0 V $V_{ds}$ from 1 GHz to 40 GHz. $S_{11}$ and $S_{22}$ are plotted in the Smith chart, while $S_{21}$ is shown in the superimposed polar plot. The measured results are scattered in black and compared with simulations plotted in a solid red line based on the developed device model.

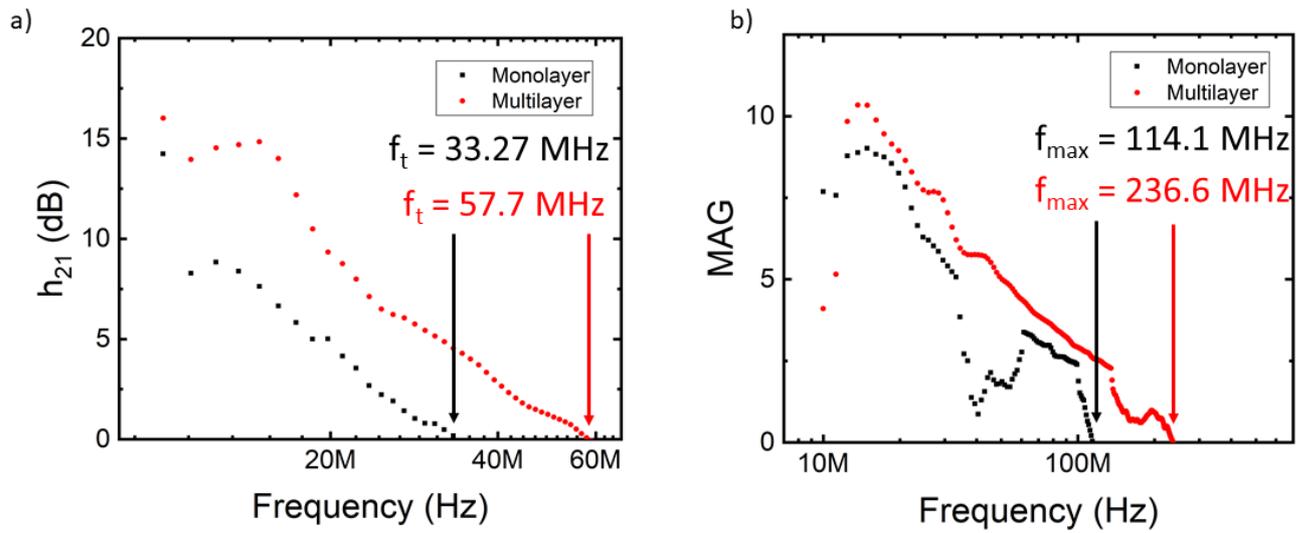

**Figure 4.** Comparison of a) the current gain, $h_{2,1}$, of the two fabricated transistors, and b) of the maximum available gain for the two types of transistors.

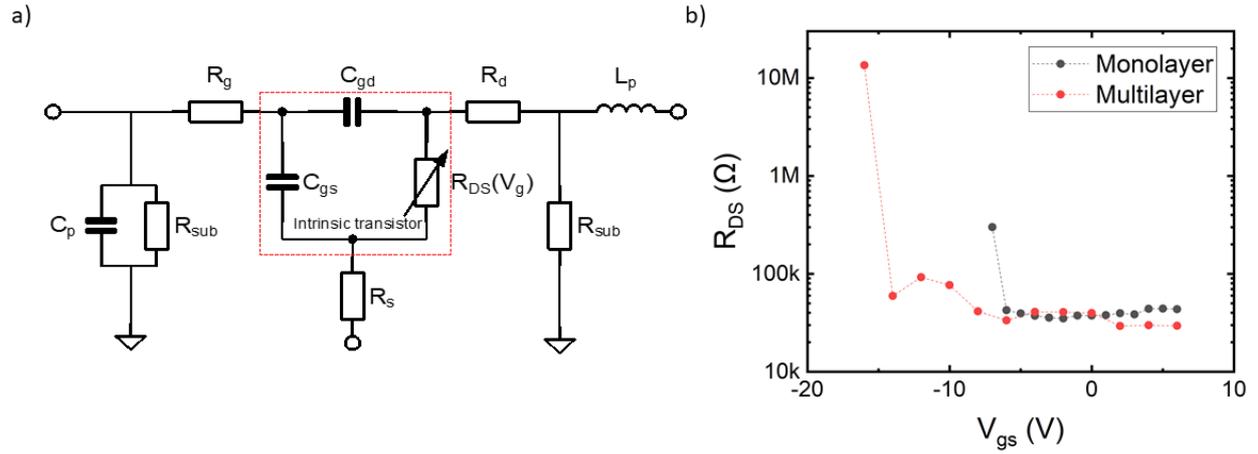

**Figure 5.** a) Small-signal equivalent circuit for the operating point at $V_{ds} = V_{gs} = 0$ V. The intrinsic components of the transistor are located inside the red dashed square. The variable $R_{DS}$ represents the nonlinearity of the channel as a function of $V_{gs}$. $R_g$, $R_d$, $R_s$ are the extrinsic elements and $C_p$, $L_p$ and $R_{sub}$ are parasitic components. b) Extracted values for $R_{DS}$ over the applied gate voltage range.

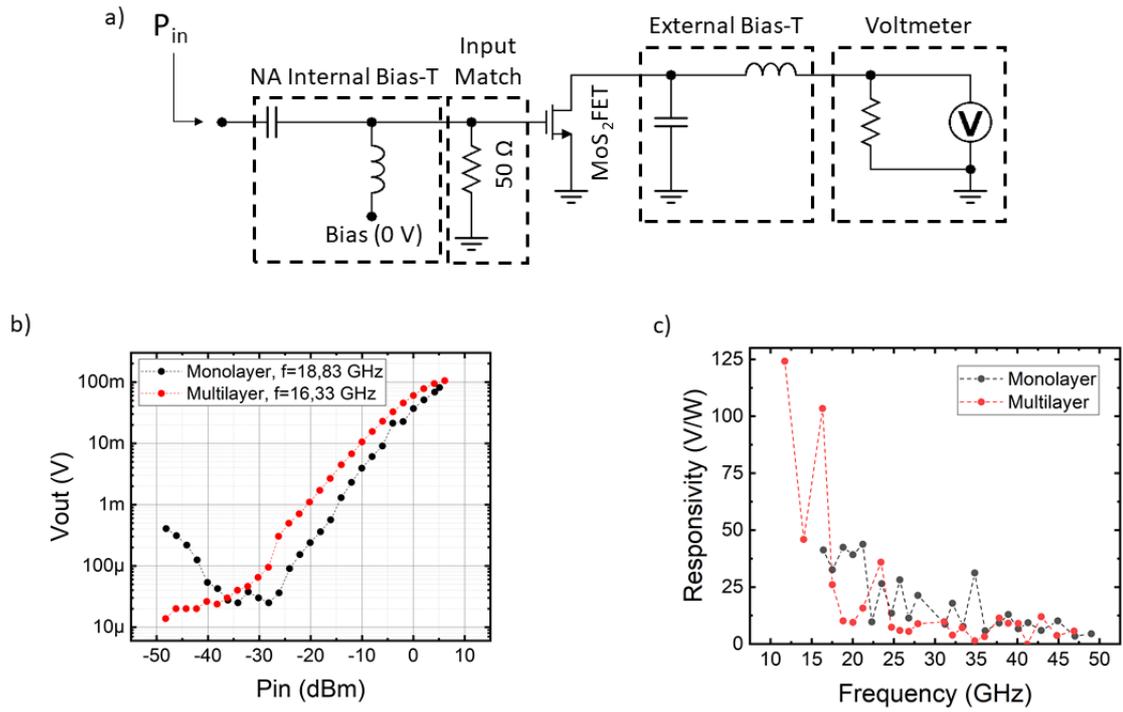

**Figure 6.** a) Schematic of the power detector with the on-wafer transistor. b) Measured output voltage as a function of the input power at 16.33 GHz for the multilayer (M) device and at 18.83 GHz for the monolayer (S) device. Both detectors show a similar dynamic range of around 30 dB. c) Responsivity comparison between the monolayer (S) and multilayer (M) devices over frequency.

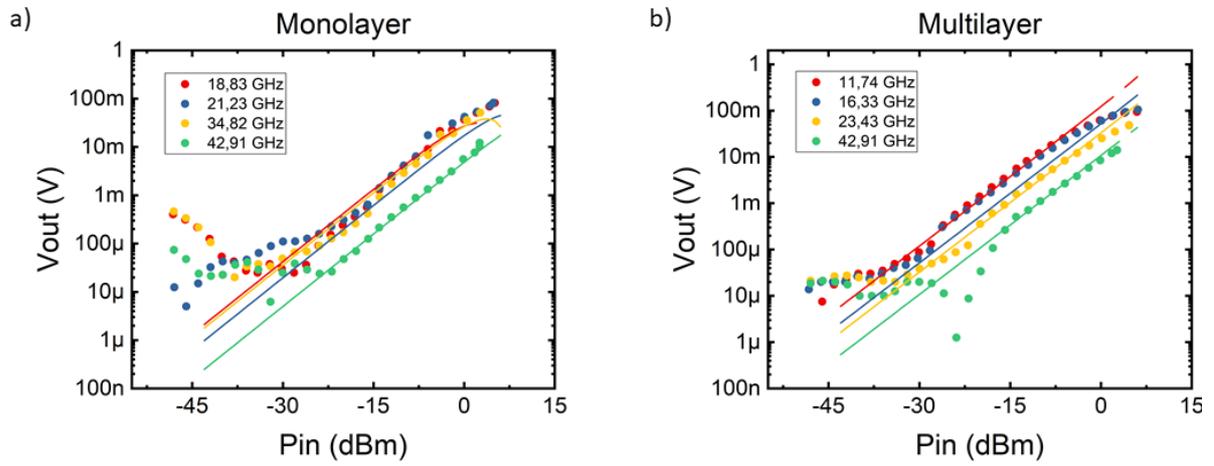

**Figure 7.** Performance comparison of the model-based simulation (solid lines) and measurement results (discrete points) in the a) monolayer (S) and b) the multilayer (M) devices.

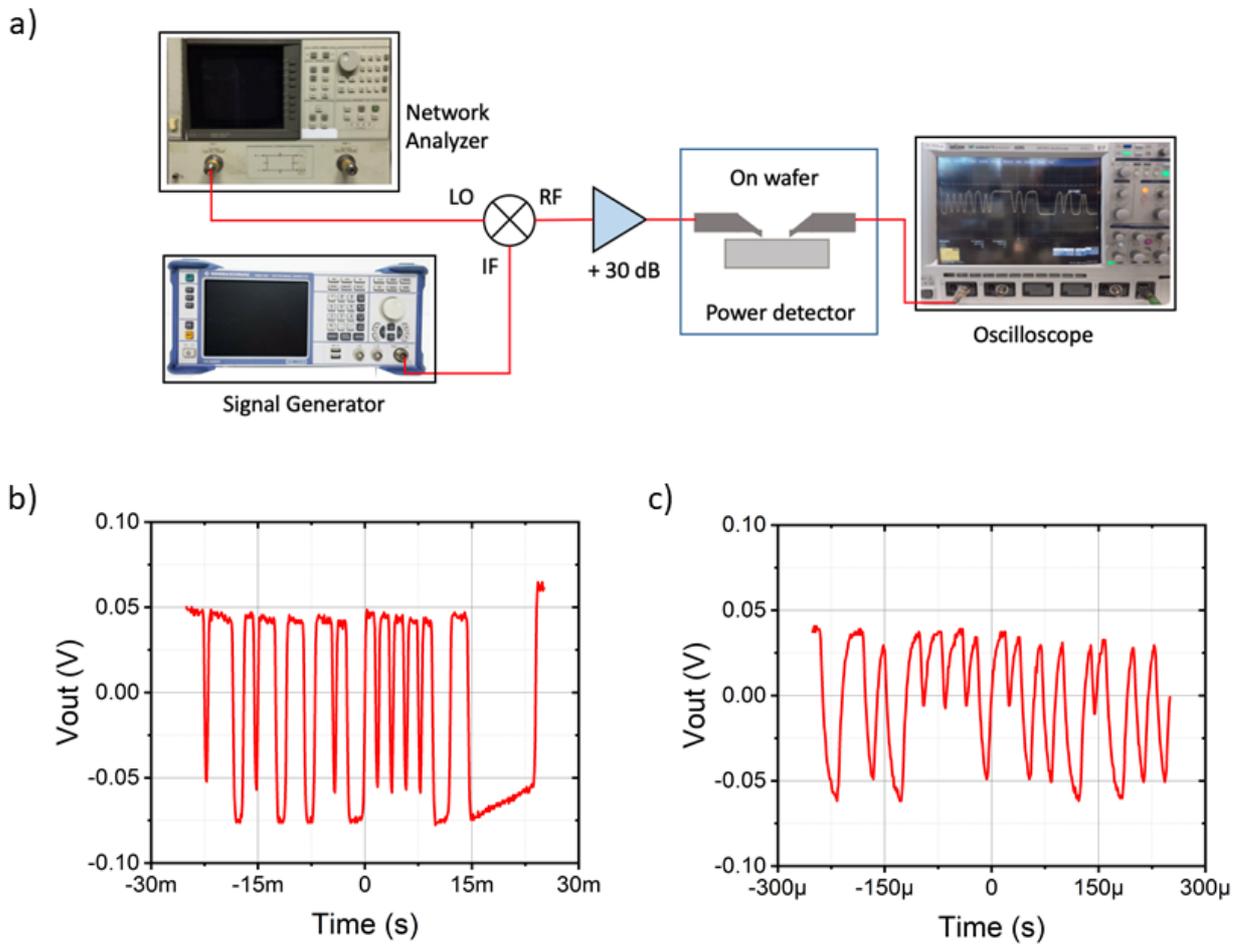

**Figure 8.** a) Block diagram of the OOK demodulator test setup for the multilayer material-based detector. The waveform response is obtained at the oscilloscope for b) 1 kSymb/s and c) 100 kSymb/s, with the high and low values representing the ON- and OFF-states, respectively.

**Table 1.** Comparison of the DC characteristics of the monolayer and multilayer material devices.

|  | $\mu_{gm}$ (cm$^2$/V·s) | $\mu_{0,\ Y\ Function}$ (cm$^2$/V·s) | $R_C$ (kΩ·μm) | Threshold Voltage (V) | Subthreshold Swing (mV/dec) | ON/OFF current ratio |
|---|---|---|---|---|---|---|
| Monolayer | 1.8 | 2.18 | 100 | -7.5 | 300 | 1.25·10$^8$ |
| Multilayer | 1.5 | 1.63 | 131 | -4.4 | 900 | 2.6·10$^4$ |

**Table 2.** Extracted values of the lumped components of the small-signal equivalent circuit

|  | $R_g$ (Ω) | $R_s$ (Ω) | $R_d$ (Ω) | $C_{gd}$ (pF) | $C_{gs}$ (pF) | $C_p$ (pF) | $L_p$ (nH) | $R_{sub1}$ (Ω) | $R_{sub2}$ (Ω) |
|---|---|---|---|---|---|---|---|---|---|
| Monolayer | 7 | 5 | 5 | 0.11 | 0.065 | 0.055 | 0.01 | 2000 | 6000 |
| Multilayer | 5 | 10 | 7 | 0.14 | 0.09 | 0.065 | 0.02 | 2000 | 6000 |

5.

**Table 3.** Comparison of the power detector in this work with the state of the art of power detectors based on bulk semiconductors and other 2D materials.

| Ref. | Technology | Substrate | Dynamic Range | Power Consumption | Responsivity | Frequency |
|---|---|---|---|---|---|---|
| [63] | 130 nm CMOS | Si | 43 – 50 dB | 35.2 mW | 23.5 mV/dB | 4 – 6 GHz |
| [64] | GaAs Tunnel Diode | GaAs | 20 dB | 0 mW | 400 -1200 V/W | 15 – 35 GHz |
| [65] | GaAs Schottky | Al$_2$O$_3$ | 25 dB | 0 mW | 6000 - 1000 V/W | 90 – 110 GHz |
| [60] | GFET | SiC | 45 dB | 0 mW | 71 – 33 V/W | 2 – 110 GHz |
| [15] | MIG diode | Si | 70 dB | 0 mW | 2.8 – 1.1 V/W | 2 – 50 GHz |
| This work | Single-layer MoS2 FET | PI | 30 dB | 0 mW | 104 V/W | 16 GHz |
|  | Multi-layer MoS2 FET |  | 30 dB | 0 mW | 45 V/W | 18 GHz |

6.

**Acknowledgements**

Funding by the German Research Foundation (DFG) through the grants MOSTFLEX, (407080863), 2D MOCVD (414268710) and ULTIMOS2 (412113712), by the European Union's Horizon 2020 research and innovation programme under the grant agreements 2D-EPL (952792), Graphene Flagship (881603), QUEFORMAL (829035), by the German Ministry of Education and Research (BMBF) through the grants NEUROTEC 2 (16ME0399, 16ME0400) and NeuroSys (03ZU1106) and by the Swiss CCMX Materials Challenge grant "Large area growth of 2D materials for device integration" is gratefully acknowledged.

**References**

[1]	B. Radisavljevic, A. Radenovic, J. Brivio, V. Giacometti, A. Kis, Nat. Nanotechnol. 2011, 6, 147.

[2]	D. Lembke, A. Kis, ACS Nano 2012, 6, 10070.

[3]	D. Akinwande, N. Petrone, J. Hone, Nat. Commun. 2014, 5, 5678.

[4]	D. Krasnozhon, D. Lembke, C. Nyffeler, Y. Leblebici, A. Kis, Nano Lett. 2014, 14, 5905.

[5]	A. Sanne, R. Ghosh, A. Rai, M. N. Yogeesh, S. H. Shin, A. Sharma, K. Jarvis, L. Mathew, R. Rao, D. Akinwande, S. Banerjee, Nano Lett. 2015, 15, 5039.

[6]	H.-Y. Chang, M. N. Yogeesh, R. Ghosh, A. Rai, A. Sanne, S. Yang, N. Lu, S. K. Banerjee, D. Akinwande, Adv. Mater. 2016, 28, 1818.

[7]	Z. Guo, S. Chen, Z. Wang, Z. Yang, F. Liu, Y. Xu, J. Wang, Y. Yi, H. Zhang, L. Liao, P. K. Chu, X.-F. Yu, Adv. Mater. 2017, 29, 1703811.

[8]	A. Hamed, O. Habibpour, M. Saeed, H. Zirath, R. Negra, IEEE Microw. Wirel. Compon. Lett. 2018, 28, 347.

[9]	K. S. Novoselov, A. K. Geim, S. V. Morozov, D. Jiang, Y. Zhang, S. V. Dubonos, I. V. Grigorieva, A. A. Firsov, Science 2004, 306, 666.

[10]	M. C. Lemme, T. J. Echtermeyer, M. Baus, H. Kurz, IEEE Electron Device Lett. 2007, 28, 282.

[11]	J. S. Moon, D. Curtis, M. Hu, D. Wong, C. McGuire, P. M. Campbell, G. Jernigan, J. L. Tedesco, B. VanMil, R. Myers-Ward, C. Eddy, D. K. Gaskill, Electron Device Lett. IEEE 2009, 30, 650.

[12]	H. Wang, A. Hsu, K. K. Kim, J. Kong, T. Palacios, 2010, p. 23.6.1-23.6.4.

[13]	M. N. Yogeesh, K. Parish, J. Lee, L. Tao, D. Akinwande, in Microw. Symp. IMS 2014 IEEE MTT- Int., 2014, pp. 1–4.

[14]	M. Saeed, A. Hamed, R. Negra, M. Shaygan, Z. Wang, D. Neumaier, in 2017 IEEE MTT-Int. Microw. Symp. IMS, 2017, pp. 1649–1652.

[15]	M. Shaygan, Z. Wang, M. S. Elsayed, M. Otto, G. Iannaccone, A. H. Ghareeb, G. Fiori, R. Negra, D. Neumaier, Nanoscale 2017, 9, 11944.

[16]	H. Pandey, S. Kataria, A. Gahoi, M. C. Lemme, IEEE Electron Device Lett. 2017, 38, 1747.

[17]	M. Saeed, A. Hamed, Z. Wang, M. Shaygan, D. Neumaier, R. Negra, IEEE Electron Device Lett. 2018, 39, 1104.

[18]	Z. Wang, B. Uzlu, M. Shaygan, M. Otto, M. Ribeiro, E. G. Marín, G. Iannaccone, G. Fiori, M. S. Elsayed, R. Negra, D. Neumaier, ACS Appl. Electron. Mater. 2019, 1, 945.

[19]	F. Schwierz, Nat. Nanotechnol. 2010, 5, 487.

[20]	S. Das, J. Appenzeller, IEEE Trans. Nanotechnol. 2011, 10, 1093.

[21]	F. Schwierz, Proc. IEEE 2013, 101, 1567.

[22]	M. C. Lemme, L.-J. Li, T. Palacios, F. Schwierz, MRS Bull. 2014, 39, 711.

*[23]  L. Li, Y. Yu, G. J. Ye, Q. Ge, X. Ou, H. Wu, D. Feng, X. H. Chen, Y. Zhang, Nat. Nanotechnol. 2014, 9, 372.*

*[24]  W. Zhu, S. Park, M. N. Yogeesh, K. M. McNicholas, S. R. Bank, D. Akinwande, Nano Lett. 2016, 16, 2301.*

*[25]  A. Avsar, I. J. Vera-Marun, J. Y. Tan, K. Watanabe, T. Taniguchi, A. H. Castro Neto, B. Özyilmaz, ACS Nano 2015, DOI 10.1021/acsnano.5b00289.*

*[26]  Y. Y. Illarionov, M. Waltl, G. Rzepa, J.-S. Kim, S. Kim, A. Dodabalapur, D. Akinwande, T. Grasser, ACS Nano 2016, 10, 9543.*

*[27]  S. Wachter, D. K. Polyushkin, O. Bethge, T. Mueller, Nat. Commun. 2017, 8, 14948.*

*[28]  J. Du, C. Ge, H. Riahi, E. Guo, M. He, C. Wang, G. Yang, K. Jin, Adv. Electron. Mater. 2020, 6, 1901408.*

*[29]  D. K. Polyushkin, S. Wachter, L. Mennel, M. Paur, M. Paliy, G. Iannaccone, G. Fiori, D. Neumaier, B. Canto, T. Mueller, Nat. Electron. 2020, 3, 486.*

*[30]  X. Zhang, J. Grajal, J. L. Vazquez-Roy, U. Radhakrishna, X. Wang, W. Chern, L. Zhou, Y. Lin, P.-C. Shen, X. Ji, X. Ling, A. Zubair, Y. Zhang, H. Wang, M. Dubey, J. Kong, M. Dresselhaus, T. Palacios, Nature 2019, 566, 368.*

*[31]  A. M. Askar, M. Saeed, A. Hamed, R. Negra, M. M. Adachi, Nanoscale 2021, 13, 8940.*

*[32]  Q. Gao, C. Zhang, K. Yang, X. Pan, Z. Zhang, J. Yang, Z. Yi, F. Chi, L. Liu, Micromachines 2021, 12, 451.*

*[33]  H. C. Torrey, C. A. Whitmer, Crystal Rectifiers, McGraw-Hill Book Company, 1948.*

*[34]  A. Zak, M. A. Andersson, M. Bauer, J. Matukas, A. Lisauskas, H. G. Roskos, J. Stake, Nano Lett. 2014, 14, 5834.*

*[35]  G. Ferrari, L. Fumagalli, M. Sampietro, E. Prati, M. Fanciulli, IEEE Microw. Wirel. Compon. Lett. 2005, 15, 445.*

*[36]  A. Valdes-Garcia, R. Venkatasubramanian, R. Srinivasan, J. Silva-Martinez, E. Sanchez-Sinencio, in 23rd IEEE VLSI Test Symp. VTS05, 2005, pp. 249–254.*

*[37]  D. M. Pozar, Microwave Engineering, Wiley, Hoboken, NJ, 2012.*

*[38]  M. Marx, A. Grundmann, Y.-R. Lin, D. Andrzejewski, T. Kümmell, G. Bacher, M. Heuken, H. Kalisch, A. Vescan, J. Electron. Mater. 2018, 47, 910.*

*[39]  H. Cun, M. Macha, H. Kim, K. Liu, Y. Zhao, T. LaGrange, A. Kis, A. Radenovic, Nano Res. 2019, 12, 2646.*

*[40]  S. Mignuzzi, A. J. Pollard, N. Bonini, B. Brennan, I. S. Gilmore, M. A. Pimenta, D. Richards, D. Roy, Phys. Rev. B 2015, 91, 195411.*

*[41]  Y. Y. Illarionov, T. Knobloch, M. Jech, M. Lanza, D. Akinwande, M. I. Vexler, T. Mueller, M. C. Lemme, G. Fiori, F. Schwierz, T. Grasser, Nat. Commun. 2020, 11, 1.*

*[42]  A. Piacentini, D. Schneider, M. Otto, B. Canto, Z. Wang, A. Radenovic, A. Kis, M. C. Lemme, D. Neumaier, in 2021 Device Res. Conf. DRC, 2021, pp. 1–2.*

*[43]  D. S. Schulman, A. J. Arnold, S. Das, Chem. Soc. Rev. 2018, 47, 3037.*

*[44]   Z. Cheng, Y. Yu, S. Singh, K. Price, S. G. Noyce, Y.-C. Lin, L. Cao, A. D. Franklin, Nano Lett. 2019, 19, 5077.*

*[45]   A. Allain, J. Kang, K. Banerjee, A. Kis, Nat. Mater. 2015, 14, 1195.*

*[46]   P.-C. Shen, C. Su, Y. Lin, A.-S. Chou, C.-C. Cheng, J.-H. Park, M.-H. Chiu, A.-Y. Lu, H.-L. Tang, M. M. Tavakoli, G. Pitner, X. Ji, Z. Cai, N. Mao, J. Wang, V. Tung, J. Li, J. Bokor, A. Zettl, C.-I. Wu, T. Palacios, L.-J. Li, J. Kong, Nature 2021, 593, 211.*

*[47]   S. Y. Kim, S. Park, W. Choi, Appl. Phys. Lett. 2016, 109, 152101.*

*[48]   G. Ghibaudo, Electron. Lett. 1988, 24, 543.*

*[49]   D. K. Schroder, Semiconductor Material and Device Characterization, Wiley-Interscience, New York, NY, USA, 2006.*

*[50]   T. Knobloch, B. Uzlu, Y. Y. Illarionov, Z. Wang, M. Otto, L. Filipovic, M. Waltl, D. Neumaier, M. C. Lemme, T. Grasser, arXiv:2104.08172 2021.*

*[51]   A. D. Bartolomeo, L. Genovese, F. Giubileo, L. Iemmo, G. Luongo, T. Foller, M. Schleberger, 2D Mater. 2017, 5, 015014.*

*[52]   S. McDonnell, R. Addou, C. Buie, R. M. Wallace, C. L. Hinkle, ACS Nano 2014, 8, 2880.*

*[53]   P. Bolshakov, P. Zhao, A. Azcatl, P. K. Hurley, R. M. Wallace, C. D. Young, Microelectron. Eng. 2017, 178, 190.*

*[54]   H. Bae, S. Jun, C.-K. Kim, B.-K. Ju, Y.-K. Choi, J. Phys. Appl. Phys. 2018, 51, 105102.*

*[55]   S. Bertolazzi, S. Bonacchi, G. Nan, A. Pershin, D. Beljonne, P. Samorì, Adv. Mater. 2017, 29, 1606760.*

*[56]   H. Lu, A. Kummel, J. Robertson, APL Mater. 2018, 6, 066104.*

*[57]   V. Dimitrov, J. B. Heng, K. Timp, O. Dimauro, R. Chan, J. Feng, W. Hafez, T. Sorsch, W. Mansfield, J. Miner, A. Kornblit, F. Klemens, J. Bower, R. Cirelli, E. Ferry, A. Taylor, M. Feng, G. Timp, in IEEE Int. Devices Meet. 2005 IEDM Tech. Dig., 2005, p. 4 pp. – 207.*

*[58]   S. A. Maas, IEEE Trans. Microw. Theory Tech. 1987, 35, 425.*

*[59]   G. Dambrine, A. Cappy, F. Heliodore, E. Playez, IEEE Trans. Microw. Theory Tech. 1988, 36, 1151.*

*[60]   J. S. Moon, H.-C. Seo, M. Antcliffe, S. Lin, C. McGuire, D. Le, L. O. Nyakiti, D. K. Gaskill, P. M. Campbell, K.-M. Lee, P. Asbeck, IEEE Electron Device Lett. 2012, 33, 1357.*

*[61]   J. Moon, H.-C. Seo, K.-A. Son, B. Yang, D. Wong, D. Le, C. McGuire, in 2014 IEEE MTT-Int. Microw. Symp. IMS2014, 2014, pp. 1–3.*

*[62]   O. Habibpour, Z. S. He, W. Strupinski, N. Rorsman, T. Ciuk, P. Ciepielewski, H. Zirath, IEEE Electron Device Lett. 2016, 37, 333.*

*[63]   K. Kim, Y. Kwon, IEEE Microw. Wirel. Compon. Lett. 2013, 23, 498.*

*[64]   O. Abdulwahid, S. G. Muttlak, J. Sexton, M. Missous, M. J. Kelly, in 2019 12th UK-Eur.-China Workshop Millim. Waves Terahertz Technol. UCMMT, 2019, pp. 1–3.*

*[65]   M. Hrobak, M. Sterns, M. Schramm, W. Stein, L.-P. Schmidt, in 2013 Eur. Microw. Conf., 2013, pp. 179–182.*

# Supporting Information

# Zero Bias Power Detector Circuits based on MoS$_2$ Field Effect Transistors on Wafer-Scale Flexible Substrates

*Eros Reato[1,2], Paula Palacios[3], Burkay Uzlu[1,2], Mohamed Saeed[3], Annika Grundmann[4], Zhenyu Wang[5], Daniel S. Schneider[1], Zhenxing Wang[1,\*], Michael Heuken[4,6], Holger Kalisch[4], Andrei Vescan[4], Alexandra Radenovic[5], Andras Kis[5], Daniel Neumaier,[1,7] Renato Negra[3,\*], Max C. Lemme[1,2,\*]*

[1] AMO GmbH, Advanced Microelectroncis Center Aachen (AMICA), Otto-Blumenthal-Strasse 25, 52074 Aachen, Germany

[2] Chair of Electronic Devices, RWTH Aachen University, Otto-Blumenthal-Strasse 25, 52074 Aachen, Germany

[3] Chair of High Frequency Electronics, RWTH-Aachen University, Kopernikusstraße 16, 52074 Aachen, Germany

[4] Compound Semiconductor Technology, RWTH Aachen University, Sommerfeldstrasse 18, 52074 Aachen, Germany

[5] School of Engineering, EPFL, BM 2141, Station 17, 1015 Lausanne, Switzerland

[6] AIXTRON SE, Dornkaulstrasse 2, 52134 Herzogenrath, Germany

[7] Bergische Universität Wuppertal, Lise-Meitner-Str. 13, 42119 Wuppertal

1. **Atomic Force Microscope (AFM) analysis and thickness measurements of the MoS2 layers**

The AFM measurements of the thickness of both materials confirmed the results of the Raman measurements, the thickness of the monolayer material 0.9 nm while the thickness of the multilayer material is around 5 nm. Furthermore, we can appreciate some substantial differences from the AFM scan of both materials shown in **Figure S 1a** and **b**. The scan on the multilayer material shows that many multilayer islands, with an area of a few hundred nanometers, are present in the surface area of the material. One can observe that the height of such islands can be around 3 nm from the base plane of the MoS$_2$, as shown in the last part of the single AFM scan in **Figure S 1c**, marked with an arrow.

A closer look at the surface of the Multilayer material is shown in **Figure S 2**. Here one can observe that some islands are made of up to 8 layers stacked on each other.

The same effect doesn't seem to be present in the monolayer material, where some patches and wrinkles are visible, probably deriving from the Na$_2$MoO$_4$ solution that was spin-coated on the wafer before the growth[1]. The RMS roughness of the monolayer and multilayer MoS$_2$ surface is calculated from the AFM maps in **Figure S 1b** and **Figure S 2** and it is $RMS_S$ = 865 pm and $RMS_M$ = 5.03 nm, for the monolayer and multilayer material respectively, both calculated on a surface area of 4 μm$^2$.

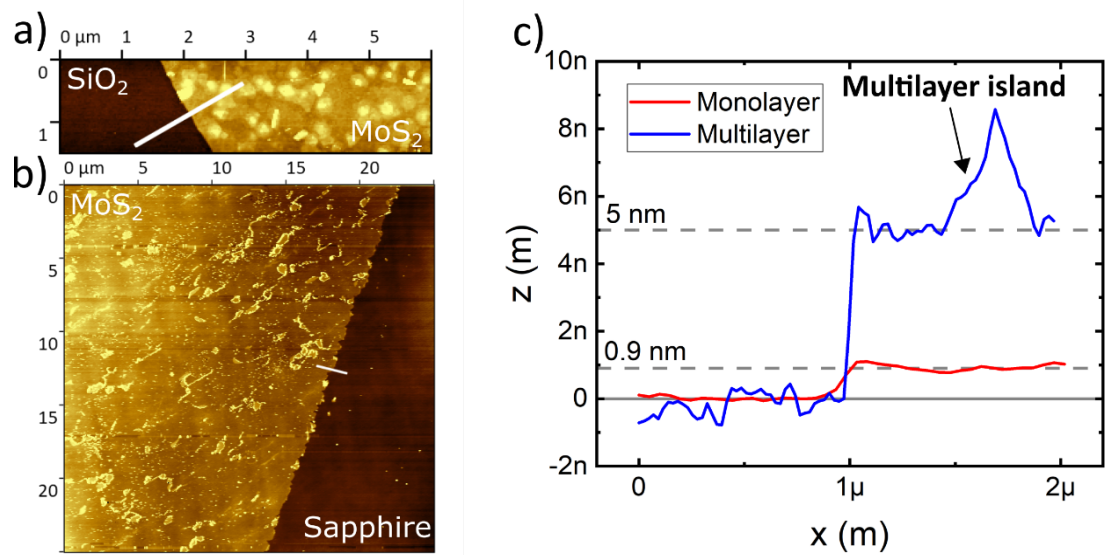

**Figure S 1.** a) AFM tapping mode scan of a multilayer MoS$_2$ sheet on a Si/SiO$_2$ substrate. b) AFM tapping mode scan of a monolayer MoS$_2$ sheet on the sapphire growth substrate. The white line represents the single line scan that was used for determining the thickness. c) Thickness measurements of both materials where zero represent the z-level of the area without MoS$_2$. The arrow indicates the thickness of one of the multilayer islands in figure a).

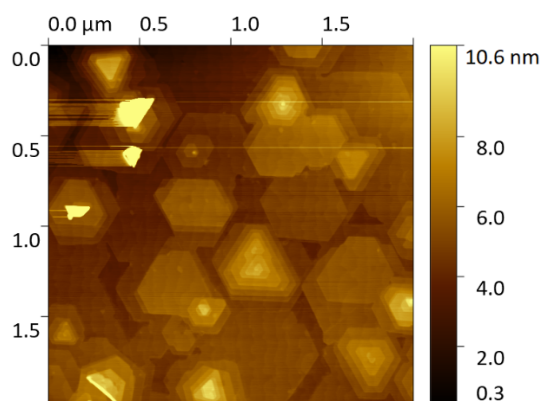

**Figure S 2.** AFM tapping mode scan of multilayer MoS$_2$ on the sapphire growth substrate.

## 2. Measurements under bending conditions

The measurements under bending conditions were performed by peeling the substrate from the carrier substrate, and by attaching the thin PI foil on top of three different bending rods with different radius, as shown in **Figure S 3a**.

The curves in **Figure S 3b** and **c** show that the peeling process alone induces some degradation of the devices since both devices seem to have around 10 times less ON current with respect to the curves before the peeling. Even the hysteresis in both cases seems to be slightly influenced.

After the peeling, the same devices were tested under different bending radii, as shown in **Figures S 3d** and **e**. For both M and S devices, the curves remain unchanged after the bending on the rod with a 25 mm radius, while the M device seems to lose some conductivity after the bending on the 12.5 mm rod. The test with a 6.2 mm rod shows that while the curve for the M device remains substantially unchanged, the S device loses most of its conductivity.

Finally, the M device used in the previous test did undergo several bending cycles. The results in **Figure S 3f** show that the M device can withstand up to 1000 bending cycles on the 6.2 mm rod without losing its characteristic.

These bending measurements show that the devices still perform well also under different bending conditions, but the S devices seem to fail when bent with the minimum bending radius. The increase of the hysteresis and the decrease of the conductivity after the mechanical peeling is due to the non-standard manual operation. During the peeling the devices could undergo a very high strain stress, possibly even more than on the bending rods, we think that this is the reason of the massive decrease of the performances. In this regard, as a future outlook, more effort needs to be put into development of a standardized peeling process that minimize the damages.

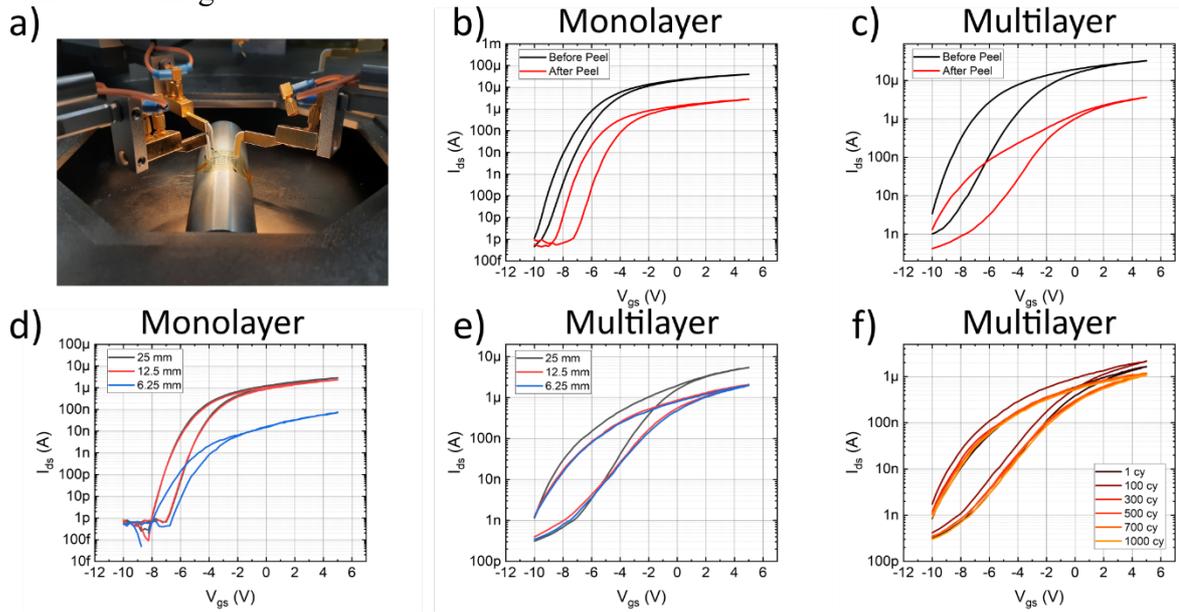

**Figure S 3.** a) Picture of the bending measurements setup. The bending rod in the picture has a radius of 12.5 mm. b) and c) Transfer characteristic of a monolayer (S) and a multilayer (M) device before and after the peeling of the PI from the substrate. d) and e) Transfer characteristic of a S and M device under 25 mm, 12.5 mm and 6.25 mm bending radiuses. f) Transfer curves of a M devices up to 1000 bending cycles.

### 3. Threshold voltage stability measurements

In order to check the stability of the threshold voltage of the devices, we performed gate voltage sweep measurements on both the M and S transistors, as seen in **Figure S 4**.
One can notice that sweeping the monolayer devices with different gate voltage ranges has almost no effect on the characteristics of the monolayer device (**Figure S 4a**), while the same operation with the multilayer devices lead to instability of the curves in terms of threshold voltage and hysteresis(**Figure S 4b**).
This effect is probably due to the different interaction between the two kind of materials and the underlying $Al_2O_3$. It seems like the whole position of the curves changes according to the gate voltage bias. A reason for that could be that a higher concentration of states in the bandgap of multilayer $MoS_2$, due to a more defective nature of the material itself and the consequent different doping levels, interacts with the defect bands in the $Al_2O_3$ layer, causing an excess of charge in the oxide during the sweeps that shifts the curves[2]. That would also explain the large difference of the $V_{TH}$ values of the M-devices in the DC and RF characterization. In this regard, more effort need to be put into the optimization of the 2D materials and their related oxides, in order to overcome these intrinsic reliability issues [3].

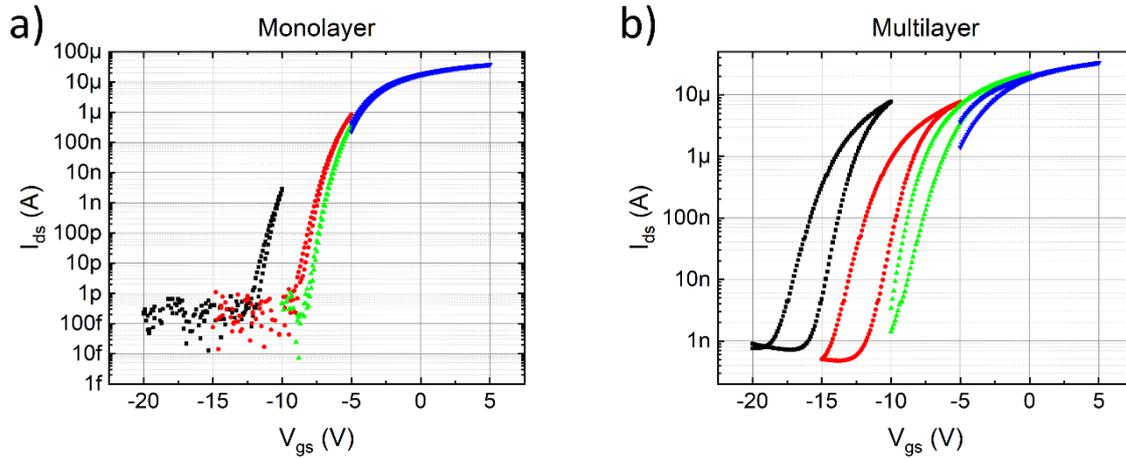

**Figure S 4.** a )and b) Transfer curve from a S and M device with different sweeping range: -20 V to -10 V, -15 V to -5 V (black), -15 V to -5 V (red), -10 V to 0 V (green), -5 V to 5 V (blue). The drain voltage is $V_{ds} = 1$ V.

**References**

[1] H. Cun, M. Macha, H. Kim, K. Liu, Y. Zhao, T. LaGrange, A. Kis, A. Radenovic, *Nano Res.* **2019**, *12*, 2646.
[2] T. Knobloch, B. Uzlu, Y. Y. Illarionov, Z. Wang, M. Otto, L. Filipovic, M. Waltl, D. Neumaier, M. C. Lemme, T. Grasser, *arXiv:2104.08172* **2021**.
[3] Y. Y. Illarionov, T. Knobloch, M. Jech, M. Lanza, D. Akinwande, M. I. Vexler, T. Mueller, M. C. Lemme, G. Fiori, F. Schwierz, T. Grasser, *Nat. Commun.* **2020**, *11*, 1.

# Zero Bias Power Detector Circuits based on MoS$_2$ Field Effect Transistors on Wafer-scale Flexible Substrates

Power detector circuits are realized based on MoS$_2$ channel transistors fabricated on flexible PI substrates. The work compares the DC and RF performance of multilayer and monolayer wafer-scale grown MoS$_2$ for power detection and it shows that both materials allow the fabrication of circuits that work at GHz frequencies with excellent dynamic range and responsivity.

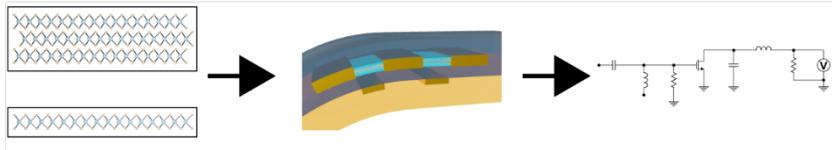